\documentclass[12pt]{article}
\newcommand{\beq}{\begin{equation}}
\newcommand{\eeq}{\end{equation}}
\newcommand{\beqa}{\begin{eqnarray}}
\newcommand{\eeqa}{\end{eqnarray}}
\newcommand{\beqar}{\begin{eqnarray*}}
\newcommand{\eeqar}{\end{eqnarray*}}

\newcommand{\eps}{\epsilon}
\newcommand{\ga}{\gamma}

\newcommand{\inn}{\!\cdot\!}

\newcommand{\la}{\lambda}

\newcommand{\z}{\zeta}

\newcommand{\eg}{{\it e.g.,}\ }
\newcommand{\ie}{{\it i.e.,}\ }
\newcommand{\labell}[1]{\label{#1}} 
\newcommand{\reef}[1]{(\ref{#1})}
\newcommand\prt{\partial}
\newcommand\veps{\varepsilon}

\newcommand\cF{{\cal F}}

\newcommand\cN{{\cal N}}

\newcommand\cA{{\cal A}}

\newcommand\cM{{\cal M}}

\newcommand\cP{{\cal P}}

\newcommand\bF{{\bf F}}
\newcommand\cB{{\cal B}}

\newcommand\bz{\bar{z}}

\newcommand\Tr{{\rm Tr}}
\newcommand\tr{{\rm tr}}
\newcommand\STr{{\rm STr}}

\begin{document}

\vspace*{1cm}

\begin{center}
{\bf \Large   S-duality of color-ordered amplitudes }

\vspace*{1cm}

{Mohammad R. Garousi\footnote{garousi@ferdowsi.um.ac.ir} }\\
\vspace*{1cm}
{ Department of Physics, Ferdowsi University of Mashhad,\\ P.O. Box 1436, Mashhad, Iran}
\\
\vspace{2cm}

\end{center}

\begin{abstract}
\baselineskip=18pt

Recently, it has been proposed that  the S-matrix elements on  the world volume of an abelian  D$_3$-brane  are consistent with the Ward identity associated with the  S-duality.  In this paper we extend this study to the case of  multiple D$_3$-branes. We speculate  that the S-matrix elements are consistent with the S-dual Ward identity irrespective of the ordering of the external states.    Imposing this symmetry on  the particular case of the S-matrix element of one Kalb-Ramond, one transverse scalar and two  nonabelian gauge  bosons,  we will find the linear S-duality transformation of the commutator of two nonabelian gauge field strengths.  Using this transformation and the standard S-duality transformations of the supergravity fields,   all other  nonabelian  S-matrix elements of one closed and three open string states can be found  by the S-duality proposal. We will show that the predicted S-matrix elements are reproduced exactly by  explicit calculations.

\end{abstract}

\vfill
\setcounter{page}{0}
\setcounter{footnote}{0}
\newpage


\section{Introduction } \label{intro}
It is known that the  type II superstring theory  is invariant under  T-duality \cite{Kikkawa:1984cp,TB,Giveon:1994fu,Alvarez:1994dn,Becker:2007zj}  and  S-duality \cite{Font:1990gx,Sen:1994fa,Rey:1989xj,Sen:1994yi,Schwarz:1993cr,Hull:1994ys,Becker:2007zj}. At the classical level, these dualities appear in  equations of motion and in their solutions \cite{Hassan:1991mq,Cvetic:1995bj,Breckenridge:1996tt,Costa:1996zd}. At the quantum level, these dualities should appear in the S-matrix elements. The  contact terms of the sphere-level S-matrix elements of four gravitons are speculated to be invariant under the S-duality after including the loops and nonperturbative effects \cite{Green:1997tv} - \cite{Basu:2007ck}. This idea has been  extended to the contact terms of the S-matrix element of two gravitons on the world volume of D$_3$-branes as well \cite{ Bachas:1999um,Basu:2008gt}. 

For other S-matrix elements, one expects they satisfy the Ward identity associated with the global S-duality \cite{Garousi:2011fc,Garousi:2011we}. The Ward identity relates a specific $n$-point function  to the other  $n$-point functions which are the transformation of  the original one under the  linear S-duality. The linear S-duality transformation is a transformation which is linear on the external states and is nonlinear  on the background fields.  Using this  Ward identity,   the S-matrix elements on the world volume of F$_1$-string/NS$_5$-brane have been proposed to be related to the corresponding  S-matrix elements on the world volume of D$_1$-string/D$_5$-brane  \cite{Garousi:2011we}.

A D$_3$-brane is invariant under the S-duality. On the other hand,  the magnetic and the electric components of the gauge boson on the world volume of the D$_3$-brane rotate into each other under the linear S-duality \cite{Gibbons:1995ap}. Therefore, the S-dual Ward identity predicts that the  S-matrix elements of $n$ gauge bosons  to be invariant under the linear S-duality transformations.  It has been shown in \cite{Garousi:2011vs} that  the  disk-level S-matrix element of four gauge bosons on the world volume of a single D$_3$-brane is invariant under the linear S-duality transformations on the external states. However, this amplitude along does not satisfy the second criteria  of the S-dual Ward identity, \ie it is not invariant under the nonlinear S-duality transformation on the background  dilaton field.  The leading $\alpha'$ order terms are invariant under the nonlinear S-duality, whereas, the non-leading order terms   have some extra dilaton factors which are not invariant.  They  must  be corrected by the loop and the nonperturbative effects \cite{Green:1997tv}. For example, the $\alpha'^2$  order terms have the dilaton factor $e^{-\phi_0}$ \cite{Garousi:2011vs} which may be extended into the $SL(2,Z)$ invariant nonholomorphic Eisenstein series $E_1(\phi_0,C_0)$ after including the loops and the nonperturbative effects. All other S-matrix elements on the world volume of  the  D$_3$-brane should satisfy the above Ward identity.  The S-dual Ward identity of some S-matrix elements   including only  closed string  have been studied in \cite{Garousi:2011we}, and   including both open and closed strings  have been studied in \cite{Garousi:2011vs}.

In this paper, we would like to study the S-dual Ward identity   of multiple D$_3$-branes. The S-matrix elements in this case should include a Chen-Paton factor for any ordering of the external states. We speculate that  the S-matrix elements for any ordering of the external states satisfy  the S-dual Ward identity. In other words, the S-matrix elements for any ordering of the external states are speculated to  be singlet under the linear $SL(2,Z)$ transformation. This Ward identity may be used to find the transformation of the external states under the linear $SL(2,R)$ transformations. The  linear $SL(2,R)$ transformation of the abelian gauge field strength has been found in \cite{Gibbons:1995ap}. The gauge field strength and its Hodge dual transforms as doublet in this transformation.   We will use  the S-dual Ward identity   to find the transformation of the nonabelian gauge field strengths and the transformation of the commutator of them. 

The   S-matrix elements on the world volume of multiple D$_3$-branes  can be arranged into two classes. One class includes the S-matrix elements which are nonzero in the abelian limit, and the other one includes the S-matrix elements which are zero in the abelian limit. The Ward identity of the first class indicates that the S-matrix elements must be a singlet combination of the abelian doublet \cite{Gibbons:1995ap}, \eg the S-matrix element of four gauge bosons have been written in \cite{Garousi:2011vs} as the singlet combination of four abelian doublets.  The Ward identity in the second class, however, indicates that the S-matrix elements must be the singlet combinations of  the doublets which are zero in the abelian case, \ie the doublets must include the nonabelian   gauge field strength and its Hodge dual and/or the commutator of gauge field strengths and its Hodge dual.  The Ward identity in the second class then  gives information about the transformation of the nonabelian gauge field strength and its commutators.  The  simplest example   is the disk-level S-matrix element of three gauge bosons \cite{Polchinski:1998rr}.  The Ward identity can be used to find the linear S-duality transformation of the  nonabelian gauge field strength.    
Another S-matrix element in the second class is the S-matrix element of one closed and
three open string states. The S-dual Ward identity  in this case  can be used to find the linear S-duality transformation of the commutator of two gauge field strengths.

The outline of the paper is as follows: In section 2, we study   the disk-level S-matrix element of three   gauge bosons  on the world volume of N $D_3$-branes. Imposing the S-dual Ward identity on this amplitude, we propose a linear S-duality transformation for the nonabelian gauge field strength. In section 3.1, we calculate the nonabelian S-matrix element of one Kalb-Ramond, one transverse scalar and two gauge bosons. We then show that the S-dual Ward identity of this amplitude  dictates that the commutator of two gauge field strengths and its Hodge dual transform as doublet under the linear $SL(2,R)$ transformations. Moreover, the Ward identity predicts two  S-matrix elements. One is the S-matrix element  of one RR 2-form, one transverse scalar and two gauge bosons, and the other one is the S-matrix element of one RR scalar, one Kalb-Ramond, one transverse scalar and two gauge bosons. In the latter amplitude the RR scalar must be constant. In section 3.2, using the doublets found in section 3.1, we show that the S-dual Ward identity predicts also the nonabelian S-matrix element of one graviton and three gauge bosons. We calculate this amplitude explicitly and find exact agreement with the S-duality result. In section 3.3, we calculate the nonabelian S-matrix element of one dilaton and three gauge bosons. Using the doublets found in section 3.1, we will show that the S-dual Ward identity predicts the S-matrix element of one RR scalar and three gauge bosons. In section 3.4, we calculate explicitly the   S-matrix element of one RR vertex operator and three open string states and find exact agreement with the amplitudes predicted by the S-duality.

\section{Three gauge bosons  amplitude}

In this section we consider the S-matrix elements of three gauge field  vertex operators in the type IIB superstring theory. This amplitude  for $123$-ordering of the external states in the Einstein frame is given by (see \eg \cite{Polchinski:1998rr}):
\beqa
A\sim iT_3e^{-\phi_0}\bigg(\z_1 \inn k_{23}\z_2 \inn\z_3 +\z_2 \inn k_{31}\z_1 \inn\z_3 +\z_3 \inn k_{12}\z_1 \inn\z_2 \bigg)\Tr(\la_1\la_2\la_3)\delta^{4}(k_1^a+k_2^a+k_3^a)\nonumber
\eeqa
where $\phi_0$ is the constant dilaton background, $k_{nm}=k_n-k_m$ for $n,m=1,2,3$,  and the vectors $\z_n^{a}$ are the polarizations of the nonabelian gauge bosons\footnote{Our index convention is that the Greek letters  $(\mu,\nu,\cdots)$ are  the indices of the space-time coordinates, the Latin letters $(a,d,c,\cdots)$ are the world-volume indices,  the letters $(i,j,k,\cdots)$ are the normal bundle indices and the letters $n,m,\cdots$ are the particle labels.}. The generators of the $U(N)$ group are given by $\la$s. The kinematic factor is antisymmetric under interchanging the particle labels, so the total amplitude which includes the $123$-ordering   and $132$-ordering   is nonzero   when the generators belong to the subgroup $SU(N)$. The  dilaton factor comes from changing the string frame metric $\eta_{ab}^S$ to the Einstein frame metric $\eta_{ab}^E$ as $\eta_{ab}^S=e^{\phi_0/2}\eta_{ab}^E$. Using the notation  $F_{nab}=i(k_{na}\z_{nb}-k_{nb}\z_{na})$, this amplitude can be rewritten as 
\beqa
A \sim T_3e^{-\phi_0}\bigg(F_{1ab}\z_2^{[b}\z_3^{a]}+F_{2ab}\z_3^{[b}\z_1^{a]}+F_{3ab}\z_1^{[b}\z_2^{a]}\bigg)\Tr(\la_1\la_2\la_3)\delta^{4}(k_1^a+k_2^a+k_3^a)\labell{3A}
\eeqa
 The amplitude satisfies the Ward identity associated with the  gauge transformation, \eg if one replaces the polarization vector $\z_1^{a}$ with its linear transformations $\z_1^{a}+k_1^a\chi $ where $\chi $ is an arbitrary scalar, the amplitude remains unchanged. The above amplitude is reproduced by the three gauge field coupling of the $SU(N)$ Yang-Mills theory.

The type IIB superstring theory is invariant under the S-duality, hence, one expects its S-matrix elements to satisfy the Ward identity corresponding to this symmetry as well, \ie if one replaces a polarization tensor with its linear S-dual polarization, the amplitude should remain unchanged up to some background dilaton factor. The dilaton factor, on the other hand, may be extended to  $SL(2,Z)$ invariant function after including the loops and nonperturbative corrections \cite{Green:1997tv} - \cite{Basu:2007ck}.  The three point function \reef{3A} receives no loop corrections \cite{Green:1981ya}, hence, the linear transformation of the gauge field polarizations in \reef{3A}, \eg $F_{1ab}$ and $\z_2^{[b}\z_3^{a]}$, should be  such  that they cancel the nontrivial transformation of the dilaton factor $e^{-\phi_0}$.  

The linear $SL(2,R)$ transformation of the $U(1)$ gauge field strength has been found in \cite{Gibbons:1995ap}. It is given by
\beqa
{\cal F}\equiv\pmatrix{*F \cr 
e^{-\phi_0}F-C_0(*F)}\longrightarrow (\Lambda^{-1})^T \pmatrix{*F \cr 
e^{-\phi_0}F-C_0(*F)}\,\,\,;\,\,\,\Lambda\in SL(2,R)\labell{F2}
\eeqa
where $(*F)^{ab}=\eps^{abcd}(F)_{cd}/2$, $F_{ab}=\prt_aA_b-\prt_bA_a$ and  $C_0$ is the constant axion background.  This transformation  is compatible with the gauge transformation because the gauge invariant object transforms to another gauge invariant object. The above transformation rule may  be used for the linear transformation of $F_{1ab}$ in \reef{3A}. However, to find the linear S-duality transformation of $\z_2^{[b}\z_3^{a]}$ we have to extend the above transformation to the nonabelian cases in which the commutator of gauge field is nonzero. For the special case that $\phi_0=C_0=0$ and $\Lambda={\rm off}$-${\rm diag}(-1,1)$, the above  S-duality is the  electric-magnetic duality 
\beqa
 F&\longrightarrow &*F\nonumber\\
 **F&\longrightarrow &- F\labell{abel}
\eeqa

We now propose an extension of this transformation to the nonabelian case. The S-duality  should be compatible  with the following  facts:

1-It should be reduced to the above transformation for the abelian case.

2-It should keep the gauge group $U(N)$ unchanged.

3-It should be consistent with the nonabelian gauge transformation.

4-It should render the 3-point function \reef{3A} to be invariant under the linear S-duality.

To define the appropriate non-abelian S-duality, we first modify the Hodge dual operator to the nonabelian case. We define the Hodge dual of the nonabelian field strength $\bF_{ab}=\prt_a A_b-\prt_b A_a+i[A_a,A_b]$ in which $A_a=A_a^{\alpha}\lambda^{\alpha}$ and   $[\lambda^{\alpha},\lambda^{\beta}]=if^{\alpha\beta\gamma}\lambda^{\gamma}$, to be
\beqa
(*'\bF)_{ab}&=&-\frac{1}{2}\eps_{abcd}(\prt^c A'^d-\prt^d A'^c+i[A'^c,A'^d])\labell{hodge}
\eeqa
where  $A'_a=A_a^{\alpha}\lambda'^{\alpha}$ and the generators satisfy the algebra $[\lambda'^{\alpha},\lambda'^{\beta}]=-if^{\alpha\beta\gamma}\lambda'^{\gamma}$.  For abelian case  $\lambda'^{\alpha}=\lambda^{\alpha}=-1$ and for  nonabelian case $\lambda'^{\alpha}=-\lambda^{\alpha}$. Note that the $*'$ operator on the linear part of the gauge field strength is the usual Hodge dual operator. 

The  transformation of the nonabelian gauge field strength $\bF$  under the gauge transformation $\delta A^{\alpha}_{a}=\prt_{a}\chi +[A_a,\chi]$ is 
\beqa
\delta\bF=[\bF,\chi]
\eeqa
The transformation of Hodge dual of $\bF$ under the gauge transformation $\delta A'_{a}=\prt_{a}\chi' +[A_a',\chi']$ is 
\beqa
\delta (*'\bF)=[*'\bF,\chi']
\eeqa
Note that both $[\lambda^{\alpha},\lambda^{\beta}]=if^{\alpha\beta\gamma}\lambda^{\gamma}$ and $[\lambda'^{\alpha},\lambda'^{\beta}]=-if^{\alpha\beta\gamma}\lambda'^{\gamma}$ are  the algebra of the $U(N)$ group.

We now use this nonabelian Hodge dual operator to define the linear   S-duality transformation of the nonabelian gauge field strength. For the special case that $\phi_0=C_0=0$ and $\Lambda={\rm off}$-${\rm diag}(-1,1)$, we propose the following linear transformation:   
\beqa
 \bF&\longrightarrow & *'\bF\labell{trans.2}\\
  **'\bF&\longrightarrow &- \bF\nonumber
\eeqa
 Obviously, the above transformation reduces    to the abelian transformation \reef{abel}, it does not change the gauge group $U(N)$ and it transforms adjoint representation to adjoint representation. Note that if one uses the symmetric trace operator, then the above transformation reduces to  $\STr F\rightarrow \STr (*F)$ which is consistent with the linear part of the nonlinear S-duality transformations proposed in \cite{Taylor:1999pr}. 
 We will show shortly that the transformation \reef{trans.2}  renders the 3-point function \reef{3A} to be invariant under the S-dual Ward identity. 
 
 The extension of the transformation \reef{trans.2} to the case that $\phi_0$ and  $C_0$ are nonzero and $\Lambda$ is arbitrary, is  given by the following doublet:
\beqa
{\cal D}&\equiv&\pmatrix{*'\bF \cr 
e^{-\phi_0}\bF-C_0(*'\bF)}\rightarrow (\Lambda^{-1})^T \pmatrix{*'\bF \cr 
e^{-\phi_0}\bF-C_0(*'\bF)}\labell{F0}
\eeqa
Using this  doublet  and the following matrix:
\beqa
 {\cal M}_0=e^{\phi_0}\pmatrix{C_0^2+e^{-2\phi_0}&C_0 \cr 
C_0&1}\labell{M}
\eeqa
which transforms under the $SL(2,R)$ transformation as\footnote{Note that the matrix $\cM$ here is the inverse of the matrix $\cM$ in \cite{Gibbons:1995ap}.}
\beqa
{\cal M}_0\rightarrow \Lambda {\cal M}_0\Lambda ^T\,,
\eeqa
one finds the following coupling is invariant under the linear S-duality:
\beqa
({\cal D}^T)_a{}^b\cM_0({\cal D})^a{}_{b}=e^{-\phi_0}\left((*'\bF)_a{}^b(*'\bF)^a{}_{b}+\bF_{a}{}^b\bF^a{}_{b}\frac{}{}\right)\labell{FA11}
\eeqa
The nonabelian gauge invariant and S-duality invariant  action is then
\beqa
S&\sim&T_3\int d^4 x \,\Tr\left(\frac{}{}({\cal D}^T)_a{}^b\cM_0({\cal D})^a{}_{b}\right)=4iT_3e^{-\phi_0}\int d^4x \,\Tr\left(F_{a}{}^b[A^a,A_b]\right)\labell{S}
\eeqa
where we have used the identity
\beqa
\frac{1}{2!}\eps_{a}{}^{cde}\eps^a{}_{c}{}^{fg}=-(\eta^{df}\eta^{eg}-\eta^{dg}\eta^{ef})\labell{iden0}
\eeqa
The Maxwell term is canceled in \reef{S} which is consistent with the fact that the Maxwell action is not invariant under the S-duality. The action \reef{S} is consistent with  the 3-point function \reef{3A}. The two commutator terms in \reef{S} is also canceled. This indicates that the four gauge bosons coupling in the Yang-Mills theory is not invariant under the S-duality. In fact this contact term combines with some massless poles reproduced by the coupling \reef{S} to produce the 4-point function which is invariant under the S-duality.

Under the S-dual Ward identity, an $n$-point function must transform to $n$-point function. Hence, one needs   $1\rightarrow 1$ and $2\rightarrow 2$ transformations of \reef{F0}. The  $1\rightarrow 1$ transformation is the same as abelian case \reef{F2}.
The $2\rightarrow 2$ transformation is  
\beqa
\cA\equiv\pmatrix{-*[A,A] \cr 
e^{-\phi_0}[A,A]+C_0(*[A,A])}\rightarrow (\Lambda^{-1})^T \pmatrix{-*[A,A] \cr 
e^{-\phi_0}[A,A]+C_0(*[A,A])}\labell{AA1}
\eeqa
where $(*[A,A])^{ab}=\eps^{abcd}[A_c,A_d]/2$. For the special case that $\phi_0=C_0=0$ and $\Lambda={\rm off}$-${\rm diag}(-1,1)$, the above  linear S-duality transformation is
\beqa
[A^a,A^b]\rightarrow -(*[A,A])^{ab}
\eeqa
For the plane waves
\beqa
A_1=\z_1\la_1e^{ik_1.x}&;&A_2=\z_2\la_2e^{ik_2.x}\labell{plane}
\eeqa
it reduces to 
\beqa
\z_1^{[a}\z_2^{b]}[\la_1,\la_2]e^{i(k_1+k_2).x}\rightarrow -*\z_1^{[a}\z_2^{b]}[\la_1,\la_2]e^{i(k_1+k_2).x}
\eeqa
which gives  the following transformation for the antisymmetric combination of two gauge field polarizations:
\beqa
\z_1^{[a}\z_2^{b]} \rightarrow -*\z_1^{[a}\z_2^{b]} 
\eeqa
The extension of this transformation   to the case that $\phi_0$ and  $C_0$ are nonzero and $\Lambda$ is arbitrary, is  then given by the following doublet:
\beqa
\cA_{12}^{ab}\equiv\pmatrix{-*\z_1^{[a}\z_2^{b]} \cr 
e^{-\phi_0}\z_1^{[a}\z_2^{b]}+C_0(*\z_1^{[a}\z_2^{b]})}\rightarrow (\Lambda^{-1})^T \pmatrix{-*\z_1^{[a}\z_2^{b]} \cr 
e^{-\phi_0}\z_1^{[a}\z_2^{b]}+C_0(*\z_1^{[a}\z_2^{b]})}\labell{AA}
\eeqa
where the subscript $12$ on $\cA$ is the particle labels.

Using the  doublets \reef{F2}, \reef{AA}  and  matrix \reef{M},
one finds the following coupling is invariant under the linear $SL(2,R)$ transformation:
\beqa
\cF^T_{1ab}\cM_0\cA_{23}^{ab}&=&e^{-\phi_0}\left(-(*F_1)_{ab}(*\z_2^{[a}\z_3^{b]})+F_{1ab} \z_2^{[a}\z_3^{b]}\frac{}{}\right)\nonumber\\
&=&2e^{-\phi_0}F_{1ab} \z_2^{[a}\z_3^{b]}\labell{FA1}
\eeqa
 where in the second line we have used the identity \reef{iden0}.
Note that if one had used the usual Hodge dual operator instead of the nonabelian Hodge dual operator \reef{hodge} in the doublet  \reef{F0}, one would find 
the right hand side of \reef{FA1} to be zero. 

The 3-point function \reef{3A} 
can then be written as 
\beqa
A\sim T_3\left(\cF^T_{1ab}\cM_0\cA_{23}^{ba}+\cF^T_{2ab}\cM_0\cA_{31}^{ba}+\cF^T_{3ab}\cM_0\cA_{12}^{ba}\right)\Tr(\la_1\la_2\la_3)\delta^{4}(k_1^a+k_2^a+k_3^a)\labell{FA}
\eeqa
which is manifestly invariant under the linear $SL(2,R)$ transformation. Note that there is no dilaton factor left over. In fact the tree-level 3-point function is the only S-matrix element which is fully invariant under the linear S-duality. In all other cases, there are some dilaton factors in the final S-dual amplitude. One needs to add the loops and the nonperturbative effects to the tree-level amplitudes to make them S-duality invariant.

Before ending this section let us compare the  Ward identity corresponding to the nonabelian gauge symmetry with the S-dual Ward identity. The nonabelian gauge symmetry of a theory indicates that the S-matrix elements in the theory must satisfy the Ward identity. In the trivial background, it is the invariance of the S-matrix elements   under the linear gauge transformation.  In the  nontrivial  background which includes the nonabelian gauge field,  the Ward identity is the invariance of the S-matrix elements under  the linear transformations on the external  states and the  full  nonlinear   transformations on the background fields. Similarly, the S-duality of a theory indicates that the S-matrix elements in the theory must satisfy the S-dual Ward identity. However, in this case the transformation of the background is always nontrivial, because the dilaton which plays the rule of the coupling constant transforms non-trivially  under the S-duality. So the S-dual Ward identity is fully satisfied only after including the loops and the nonperturbatie corrections to make the dilaton background in the tree-level S-matrix element to be invariant under the full nonlinear S-duality transformation.

The study of the S-dual Ward identity of the S-matrix elements, hence,  has two parts. One part which is easy to study, is to make  the tree-level S-matrix elements to be invariant under the linear S-duality transformation on the external states, and the other part which is  nontrivial to explore, is to make  the background dilaton field in the tree-level amplitude to be invariant under the full nonlinear S-duality. This can be done by including the loops and the nonperturbative effects \cite{Green:1997tv} - \cite{Basu:2007ck}. In the first part, a S-matrix element is either invariant by itself, \eg the 3-point function \reef{FA} or the 4-point function \cite{Garousi:2011vs}, or it has to be combined with some other tree-level S-matrix elements to become invariant under the linear S-duality. We will show in the next section that  even  the first part of the  Ward identity has nontrivial information, \ie it can be used to find the linear S-duality transformations of the commutator of nonabelian gauge field strengths, and to find new S-matrix elements from a given S-matrix element.

\section{Three open and one closed string amplitudes}

We have seen that the S-dual ward identity of the 3-point function enables  us to find the linear S-duality of the commutator of two gauge fields \reef{AA1}. However, this commutator term is not invariant under the linear gauge transformation. The on-shell conditions in the 3-point function \reef{3A} in which this commutator term appears,  are such that they make the amplitude to be invariant under the   gauge transformation. This symmetry however does not allow the higher point functions to be written in terms of the commutator of two gauge bosons. In this section, by working on some examples, we are going to show that the higher point functions may be written in terms of the linear gauge field strengths and the commutator of them. The S-dual Ward identity  can then be used to find their linear S-duality  transformations. The commutator terms appear only in the S-matrix elements which are zero in the abelian case. Moreover, the S-matrix elements which involve both closed and open string states, in general, do not satisfy the S-dual Ward identity. One needs to include some other S-matrix elements to make them invariant. In this way the S-dual Ward identity can be used to find  new tree-level S-matrix elements.

\subsection{Kalb-Ramond  amplitude}

 The simplest nonabelian S-matrix element which is zero for abelian case is the S-matrix element of one Kalb-Ramond B-field, one transverse scalar and two gauge bosons\footnote{The scattering amplitude of one closed and three open string states has been studied in \cite{Garousi:2000ea,Fotopoulos:2001pt}.}.   This amplitude is given by the following correlation function:
\beqa
A&\sim&<V_{B}^{(-1,-1)}(\veps_B,p)V_{\Phi}^{(0)}(\z_1,k_1)V_{A}^{(0)}(\z_2,k_2)V_{A}^{(0)}(\z_3,k_3)>\labell{amp21}
\eeqa
Using the doubling trick \cite{Garousi:1996ad,Hashimoto:1996bf}, the vertex operators are given by the following integrals on the upper half $z$-plane:\footnote{Our conventions set $\alpha'=2$. }
\beqa
V_{B}^{(-1,-1)}&\!\!\!\!\!=\!\!\!\!\!&(\veps_B.D)_{\mu\nu}\int d^2z:\psi^{\mu}e^{-\phi}e^{ip\cdot X}:\psi^{\nu}e^{-\phi}e^{ip\cdot D\cdot X}:\nonumber\\
V_{\Phi}^{(0)}&\!\!\!\!\!=\!\!\!\!\!&(\z_1)_{i}\int dx_1:(\prt X^{i}+2ik_1\inn\psi\psi^{i})e^{2ik_1\cdot X}:\nonumber\\
V_{A}^{(0)}&\!\!\!\!\!=\!\!\!\!\!&(\z_2)_{a_2}\int dx_2:(\prt X^{a_2}+2ik_2\inn\psi\psi^{a_2})e^{2ik_2\cdot X}:\nonumber\\
V_{A}^{(0)}&\!\!\!\!\!=\!\!\!\!\!&(\z_3)_{a_3}\int dx_3:(\prt X^{a_3}+2ik_3\inn\psi\psi^{a_3})e^{2ik_3\cdot X}:
\eeqa
where  the matrix $D^{\mu}_{\nu}$ is diagonal with $+1$ in the world volume directions and $-1$ in the transverse direction. To simplify  the calculation, we assume the polarization of the B-field is along the word volume of the D$_p$-brane. The B-field polarization $\veps_B$ is antisymmetric. The on-shell conditions are $\veps_B.p=p.\veps_B=p_{\mu}p^{\mu}=0$, $k_1.k_1=0$  and $\z_i.k_i=k_i.k_i=0$ for $i=2,3$. The metric in the inner products is the flat world-volume metric. The open string momenta are along the brane whereas the transverse scalar is orthogonal to the brane.

Using the standard world-sheet propagators
\beqa
<X^{\mu}(x)X^{\nu}(y)>&=&-\eta^{\mu\nu}\log(x-y)\nonumber\\
<\psi^{\mu}(x)\psi^{\nu}(y)>&=&-\frac{\eta^{\mu\nu}}{x-y}\nonumber\\
<\phi(x)\phi(y)>&=&-\log(x-y)\labell{wpro}
\eeqa
one can calculate the correlators in \reef{amp21}. The result is
\beqa
A&\sim&p_i\z _ 1^i\bigg[  \z _ 2.\z _ 3 k_ 2. \veps_B.k_ 3 J_1 + 
     k_ 2.\z _ 3 k_ 2.\veps_B.\z _ 2 J_1 - 
     k_ 3.\z _ 2 k_ 2.\veps_B.\z _ 3 J_1 + 
     k_ 2.\z _ 3 k_ 3.\veps_B.\z _ 2 J_1 \nonumber\\
     &&- 
     k_ 3.\z _ 2 k_ 3.\veps_B.\z _ 3 J_1 + 
     k_ 2. k_ 3\z _ 2.\veps_B.\z _ 3 J_1 + 
     k_ 1.\z _ 3 k_ 2.\veps_B.\z _ 2 J_2 + 
     k_ 1.\z _ 2 k_ 3.\veps_B.\z _ 3 J_3\bigg]\labell{amp3}
\eeqa
The amplitude has also a delta Dirac function imposing the momentum conservation along the brane, \ie
\beqa
p+p.D+2k_1+2k_2+2k_3&=&0
\eeqa 
In the amplitude \reef{amp3},  $J_1,J_2,J_3$ are the following integrals
\beqa
J_1=\frac{ \left(x_2 \left(2 x_3-z_1-\bar{z}_1\right)+2 z_1 \bar{z}_1-x_3 \left(z_1+\bar{z}_1\right)\right)}{\left(x_2-x_3\right) \left(x_1-z_1\right) \left(x_2-z_1\right) \left(x_3-z_1\right) \left(x_1-\bar{z}_1\right) \left(x_2-\bar{z}_1\right) \left(x_3-\bar{z}_1\right)}K\nonumber\\
J_2=\frac{ \left(x_1 \left(2 x_3-z_1-\bar{z}_1\right)+2 z_1 \bar{z}_1-x_3 \left(z_1+\bar{z}_1\right)\right)}{\left(x_1-x_3\right) \left(x_1-z_1\right) \left(x_2-z_1\right) \left(x_3-z_1\right) \left(x_1-\bar{z}_1\right) \left(x_2-\bar{z}_1\right) \left(x_3-\bar{z}_1\right)}K\nonumber\\
J_3=\frac{ \left(2 z_1 \bar{z}_1-x_2 \left(z_1+\bar{z}_1\right)+x_1 \left(2 x_2-z_1-\bar{z}_1\right)\right)}{\left(x_1-x_2\right) \left(x_1-z_1\right) \left(x_2-z_1\right) \left(x_3-z_1\right) \left(x_1-\bar{z}_1\right) \left(x_2-\bar{z}_1\right) \left(x_3-\bar{z}_1\right)}K
\eeqa
 There is a measure $\int d^2z_1dx_1dx_2dx_3$ for all the integrals which we have omitted.  The function $K$ is
\beqa
K&=&(z-\bz)^{p.D.p}|z-x_1| ^{2k_1.p }|z-x_2| ^{2k_2.p }|z-x_3| ^{2k_3.p }(x_{12})^{4k_1.k_2}(x_{13})^{4k_1.k_3}(x_{23})^{4k_2.k_3}\labell{K}
\eeqa
Note that the integral $J_2(J_3)$ is the same as $J_1$ in which the replacement $1\leftrightarrow 2(2\leftrightarrow 1,3\leftrightarrow 2)$ has been used. Imposing the Ward identity corresponding to the gauge transformation, one finds the following relations between the integrals:
\beqa
k_1.k_3J_2= -k_2.k_3 J_1&;&k_1.k_2J_3= k_2.k_3 J_1 \labell{rel}
\eeqa
Using these relations, one can write the amplitude \reef{amp3} in terms of gauge field strength as 
\beqa
A&\!\!\!\!\!\sim\!\!\!\!\!&e^{-3\phi_0/2}p_i\z _ 1^i\bigg[\tr(\veps_B.F_2.F_3)+\frac{1}{2k_1.k_3}\tr(\veps_B.F_2)k_1.F_3.k_2-\frac{1}{2k_1.k_2}\tr(\veps_B.F_3)k_1.F_2.k_3\bigg]J_1\labell{amp4}
\eeqa
where we have also used  the Einstein frame metric in witting the above result.   The above result is valid for any ordering of the external states. The integral $J_1$ can be evaluated after fixing the ordering of the external states. One finds that this integral for the $132$ ordering is minus of the $123$ ordering. For the $123$ ordering, one has to include the Chan-Paten factor $\Tr(\la_1\la_2\la_3)$ and for the $132$ ordering the factor $\Tr(\la_1\la_3\la_2)$. So the total amplitude which is the sum of these two ordering is symmetric under changing $2\leftrightarrow 3$ and is zero for abelian case. 

The scattering amplitude for any ordering of the external states, \eg \reef{amp4}  satisfies the Ward identity corresponding to the gauge transformation, so we expect the amplitude for the case of D$_3$-brane  satisfies the Ward identity corresponding to the S-duality as well. Unlike the gauge transformation which maps the gauge field to itself up to a derivative of a scalar field, the S-duality in general maps a field to another field so the amplitude in general should be covariant under the linear S-duality transformation. For example, as we will see in a moment the amplitude \reef{amp4}  transforms to two other amplitudes under the linear S-duality. One is the scattering amplitude of one RR two-form, one transverse scalar and two gauge fields, and the other one is the scattering amplitude of one RR scalar, one B-field, one transverse scalar and two gauge fields. In other words, the S-dual Ward identity predicts that the combination of the above three amplitudes forms a S-dual multiplet. 

To find the S-dual multiplet corresponding to the amplitude \reef{amp4}, we note that   a single transverse scalar is invariant under the S-duality\footnote{A proposal for the S-duality transformation of the antisymmetric combination of three scalar fields has been proposed in \cite{Taylor:1999pr}. }. The B-field and the RR two-form  appear as doublet under the $SL(2,R)$ transformation \cite{Green:1996qg} 
\beqa
\cB\equiv\pmatrix{B \cr 
C^{(2)}}\rightarrow (\Lambda^{-1})^T \pmatrix{B \cr 
C^{(2)}}\labell{2}
\eeqa
Using the above transformation for the B-field,  the S-dual Ward identity of the amplitude \reef{amp4} dictates that there are two doublets constructed from the gauge field strength.  To find the precise form of these doublet, let us define the  2-forms  $M $ and $M^c$ as
\beqa
M_{ab}&=&[F_{ac},F^c{}_b]\nonumber\\
M_{ab}^c&=&[\prt_dF_{ab},F^{dc}]\labell{2-form}
\eeqa
which are zero for abelian case. We propose the linear S-duality transformation of these 2-forms to be the same as $F_{ab}$ in \reef{F2}, \eg
\beqa
\cM_{ab}\equiv\pmatrix{(*M)_{ab} \cr 
e^{-\phi_0}M_{ab}-C_0(*M)_{ab}}\rightarrow (\Lambda^{-1})^T \pmatrix{(*M)_{ab} \cr 
e^{-\phi_0}M_{ab}-C_0(*M)_{ab}}
\eeqa
 Using the plane waves \reef{plane}, one  finds the following transformations for the polarization tensors:
\beqa
(\cM^{12})^{ab} &\!\!\!\!\!\equiv\!\!\!\!\!&\pmatrix{*F_1^{[a}.F_2^{b]} \cr 
e^{-\phi_0}F_1^{[a}.F_2^{b]}-C_0(*F_1^{[a}.F_2^{b]})}\rightarrow (\Lambda^{-1})^T \pmatrix{*F_1^{[a}.F_2^{b]} \cr 
e^{-\phi_0}F_1^{[a}.F_2^{b]}-C_0(*F_1^{[a}.F_2^{b]})}\labell{3}\\
(\cM^{12c})^{ab} &\!\!\!\!\!\equiv\!\!\!\!\!&\pmatrix{*F_1^{ab}k_1.F_2^{c} \cr 
e^{-\phi_0}F_1^{ab}k_1.F_2^{c}-C_0(*F_1^{ab}k_1.F_2^{c})}\rightarrow (\Lambda^{-1})^T \pmatrix{*F_1^{ab}k_1.F_2^{c} \cr 
e^{-\phi_0}F_1^{ab}k_1.F_2^{c}-C_0(*F_1^{ab}k_1.F_2^{c})}\nonumber
\eeqa
where the superscript $12$ on $\cM_{ab}$ and $\cM_{ab}^c$ is the particle labels. Using the doublets \reef{2}, \reef{3} and the matrix \reef{M}, one finds the following expressions are invariant under the linear S-duality:
\beqa
(\cB^T)^{ab}\cM_0\cM^{12}_{ab}&=&e^{-\phi_0}\veps_B^{ab}M^{12}_{ab}+*\veps_C^{ab}M^{12}_{ab} -C_0(*\veps_B^{ab})M^{12}_{ab}\\
(\cB^T)^{ab}\cM_0\cM^{12c}_{ab}&=&e^{-\phi_0}\veps_B^{ab}M^{12c}_{ab}+*\veps_C^{ab}M^{12c}_{ab}-C_0(*\veps_B^{ab})M^{12c}_{ab}\nonumber
\eeqa
where $(M^{12})^{ab}=F_1^{[a}.F_2^{b]}$, $(M^{12c})^{ab}=F_1^{ab}k_1.F_2^{c}$  and $\veps_C$ is the polarization of the RR two-form. 
The amplitude \reef{amp4} can then be extended to the following linear S-dual invariant form:
\beqa
{\cal A}&\sim&   
p_i\z _ 1^ie^{-\phi_0/2}(\cB^T)^{ab}\cM_0
 \bigg[\cM^{23}_{ab}   +\frac{1}{2k_1.k_3} \cM^{23c}_{ab}k_{1c}-\frac{1}{2k_1.k_2} \cM^{32c}_{ab}k_{1c}\bigg] J_1\labell{amp6}
\eeqa
 Note that the Mandelstam variables in the integral $J_1 $ are multiplied by the dilaton factor $e^{-\phi_0/2}$, so the leading order term of the amplitude at low energy which is reproduced by the supergravity and the $\cN=4$ Super Yang-Mills theory,  has no extra dilaton factor.

The S-dual amplitude \reef{amp6} confirms that the nonabelian 2-forms \reef{2-form} transform as doublet under the linear S-duality. This amplitude predicts that the amplitude \reef{amp4} should be combined with two other amplitudes to make a S-dual multiplet.  One of them is the S-matrix element of one RR 2-form, one scalar and two gauge fields which is given by 
\beqa
 A_{C^{(2)}\Phi AA}&\!\!\!\!\!\sim\!\!\!\!\!& 
p_i\z _ 1^ie^{-\phi_0/2} \bigg[\tr(*\veps_C.F_2.F_3)\labell{amp7}\\
&&+\frac{1}{2k_1.k_3}\tr(*\veps_C.F_2)k_1.F_3.k_2-\frac{1}{2k_1.k_2}\tr(*\veps_C.F_3)k_1.F_2.k_3\bigg]J_1 \nonumber
\eeqa
  The other one is the S-matrix element of one RR scalar, one B-field, one scalar and two gauge fields which is given by the following amplitude:
\beqa
 A_{C^{(0)}B\Phi AA}&\!\!\!\!\!\sim\!\!\!\!\!&  
-p_i\z _ 1^ie^{-\phi_0/2}C_0 \bigg[\tr(*\veps_B.F_2.F_3)\labell{amp8}\\
&&+\frac{1}{2k_1.k_3}\tr(*\veps_B.F_2)k_1.F_3.k_2-\frac{1}{2k_1.k_2}\tr(*\veps_B.F_3)k_1.F_2.k_3\bigg]J_1\nonumber
\eeqa
  In the above amplitude the RR scalar is constant. We will compare the amplitude \reef{amp7} with explicit calculation in section 4.4. In the next section we will show that the nonabelian doublets found in this section  can be used to even  predict  the  non-abelian S-matrix element of one graviton and three gauge bosons.

\subsection{Graviton amplitude}

The S-matrix element of one graviton and three gauge bosons is zero for the abelian case. So the S-dual form of the amplitude should include the nonabelian doublets that we have found in the previous section. To make a singlet from the nonabelian doublets \reef{3}, the amplitude must include another doublet. Since the graviton in the Einstein frame is invariant under the S-duality, the only possibility for the graviton amplitude which has three gauge bosons,  is then to have one abelian doublet \reef{F2}. So we have to make a nonzero singlet from the graviton, the abelian doublet \reef{F2} and the nonabelian doublets \reef{3}. There are two possibilities for the polarization of the graviton in the S-dual form of the amplitude. One is the world-volume trace of the graviton   multiplies with the singlet from the two doublets, and the other one is contraction of the graviton with one of the doublet and then making a singlet from them. In the former case, however, the singlet is zero using the identity \reef{iden0}, \ie $(\cF^T)^{ab}\cM_0\cM_{ab}=0$. Hence, the only possibility for making a nonzero singlet from the graviton, the abelian doublet \reef{F2} and the nonabelian doublets \reef{3} is the following:
\beqa
 {\cal A} &\sim& 
 e^{-\phi_0/2}   (\veps.\cF_1^T)^{[ab]}\cM_0\bigg[\cM^{23}_{ab}+\frac{1}{2k_1.k_3} \cM^{23c}_{ab}k_{1c} -\frac{1}{2k_1.k_2} \cM_{ab}^{32c}k_{1c} \bigg]J_1\nonumber\\
 &&+e^{-\phi_0/2}   (\veps.\cF_2^T)^{[ab]}\cM_0\bigg[\cM^{13}_{ab}+\frac{1}{2k_2.k_3} \cM^{13c}_{ab}k_{1c} -\frac{1}{2k_1.k_2} \cM_{ab}^{31c}k_{1c} \bigg]J_2\nonumber\\
 &&+e^{-\phi_0/2}   (\veps.\cF_3^T)^{[ab]}\cM_0\bigg[\cM^{12}_{ab}+\frac{1}{2k_2.k_3} \cM^{12c}_{ab}k_{1c} -\frac{1}{2k_1.k_3} \cM_{ab}^{21c}k_{1c} \bigg]J_3\labell{ampi}
\eeqa
Our notation is $(\veps.F_1)^{[ab]}=((\veps.F_1)^{ab}-(\veps.F_1)^{ba})/2$. The above amplitude is the prediction of the S-duality for any ordering of the external states. The integrals $J_1,\, J_2,\, J_3$ can be calculated after fixing the ordering of the external states in which we are not interested in this paper.

Now let us calculate  the S-matrix element of one graviton and three  gauge bosons explicitly.  This amplitude is given by the following correlation function:
\beqa
A&\sim&<V_{g}^{(-1,-1)}(\veps,p)V_{A}^{(0)}(\z_1,k_1)V_{A}^{(0)}(\z_2,k_2)V_{A}^{(0)}(\z_3,k_3)>\labell{amp211}
\eeqa
where the graviton vertex operator is the same as the B-field vertex operator in \reef{amp21} in which the polarization is symmetric. Using the propagators \reef{wpro} one can  perform the above correlators. The result is a lengthy expression in terms of 14 integrals. Some of them have tachyon poles. Since the total amplitude should have no tachyon, there should be some relation between such integrals  and the integrals which have no tachyon poles. These relations can be found by imposing the Ward identity corresponding to the gauge transformation. Using these relations to eliminate those integrals that have tachyon poles,  one finds the following result:
\beqa
A&\sim&\bigg[\frac{1}{4}\Tr(\veps.D)(-k_ 1.\z _ 2 k_ 2.\z _ 1 k_ 2.\z _ 3 - 
 k_ 1.\z _ 2 k_ 2.\z _ 3 k_ 3.\z _ 1 + 
 k_ 1.\z _ 3 k_ 2.\z _ 1 k_ 3.\z _ 2 + 
 k_ 1.\z _ 3 k_ 3.\z _ 1 k_ 3.\z _ 2 -\nonumber\\ && 
 k_ 1.\z _ 3 k_ 2. k_ 3\z _ 1.\z _ 2 + 
 k_ 1. k_ 2 k_ 2.\z _ 3\z _ 1.\z _ 2 + 
 k_ 1. k_ 3 k_ 2.\z _ 3\z _ 1.\z _ 2 + 
 k_ 1.\z _ 2 k_ 2. k_ 3\z _ 1.\z _ 3 - \nonumber\\ &&
 k_ 1. k_ 2 k_ 3.\z _ 2\z _ 1.\z _ 3 - 
 k_ 1. k_ 3 k_ 3.\z _ 2\z _ 1.\z _ 3 - 
 k_ 1. k_ 3 k_ 2.\z _ 1\z _ 2.\z _ 3 + 
 k_ 1. k_ 2 k_ 3.\z _ 1\z _ 2.\z _ 3) - \nonumber\\ &&
 k_ 2.\z _ 3\z _ 1.\z _ 2 k_ 1.\veps.k_ 2 + 
 k_ 3.\z _ 2\z _ 1.\z _ 3 k_ 1.\veps.k_ 2 - 
 k_ 3.\z _ 1\z _ 2.\z _ 3 k_ 1.\veps.k_ 2 - 
 k_ 2.\z _ 3\z _ 1.\z _ 2 k_ 1.\veps.k_ 3 +\nonumber\\ && 
 k_ 3.\z _ 2\z _ 1.\z _ 3 k_ 1.\veps.k_ 3 + 
 k_ 2.\z _ 1\z _ 2.\z _ 3 k_ 1.\veps.k_ 3 + 
 k_ 2.\z _ 1 k_ 2.\z _ 3 k_ 1.\veps.\z _ 2 + 
 k_ 2.\z _ 3 k_ 3.\z _ 1 k_ 1.\veps.\z _ 2 - \nonumber\\ &&
 k_ 2. k_ 3\z _ 1.\z _ 3 k_ 1.\veps.\z _ 2 - 
 k_ 2.\z _ 1 k_ 3.\z _ 2 k_ 1.\veps.\z _ 3 - 
 k_ 3.\z _ 1 k_ 3.\z _ 2 k_ 1.\veps.\z _ 3 + 
 k_ 2. k_ 3\z _ 1.\z _ 2 k_ 1.\veps.\z _ 3 + \nonumber\\ &&
 k_ 1.\z _ 2 k_ 2.\z _ 3 k_ 2.\veps.\z _ 1 - 
 k_ 1.\z _ 3 k_ 3.\z _ 2 k_ 2.\veps.\z _ 1 + 
 k_ 1. k_ 3\z _ 2.\z _ 3 k_ 2.\veps.\z _ 1 + 
 k_ 1.\z _ 2 k_ 2.\z _ 3 k_ 3.\veps.\z _ 1 -\nonumber\\ && 
 k_ 1.\z _ 3 k_ 3.\z _ 2 k_ 3.\veps.\z _ 1 - 
 k_ 1. k_ 2\z _ 2.\z _ 3 k_ 3.\veps.\z _ 1 + 
 k_ 1.\z _ 3 k_ 2. k_ 3\z _ 1.\veps.\z _ 2 - 
 k_ 1. k_ 2 k_ 2.\z _ 3\z _ 1.\veps.\z _ 2 - \nonumber\\ &&
 k_ 1. k_ 3 k_ 2.\z _ 3\z _ 1.\veps.\z _ 2 - 
 k_ 1.\z _ 2 k_ 2. k_ 3\z _ 1.\veps.\z _ 3 + 
 k_ 1. k_ 2 k_ 3.\z _ 2\z _ 1.\veps.\z _ 3 + 
 k_ 1. k_ 3 k_ 3.\z _ 2\z _ 1.\veps.\z _ 3\bigg]J_1\nonumber\\
 &&+[1\leftrightarrow 2]J_2 +[2\leftrightarrow 1,3\leftrightarrow 2]J_3\labell{am4}
\eeqa
where $J_1, J_2, J_3 $ are the same integrals that appear in the amplitude \reef{amp3}.  In above amplitude we have not used the traceless condition of the graviton, so one can use the above amplitude to find the S-matrix element of one dilaton and three gauge boson. We will study such  amplitude in the next section.  

To study the linear S-duality of the above amplitude, we have to first write  it in terms of gauge field strengths whose duality transformations have been found in the previous section.  One  can   write the above amplitude   as
\beqa
A&\!\!\!\!\!\sim\!\!\!\!\!& e^{-3\phi_0/2}[(\veps.F_1)^{[ab]}-\frac{1}{8}\Tr(\veps.D)F_{1}^{ab}]\bigg[(F_2.F_3)_{ab}+\frac{1}{2k_1.k_3} (F_{2})_{ab}k_1.F_3.k_2\nonumber\\
&&-\frac{1}{2k_1.k_2} (F_{3})_{ab}k_1.F_2.k_3\bigg]J_1+(1\leftrightarrow 2)+(2\leftrightarrow 1,3\leftrightarrow 2)\labell{hAAA}
\eeqa
where  metric is in the Einstein frame.  In witting the above result, we have used the following relations:
\beqa
\frac{1}{2k_1.k_3}k_1.F_3.k_2J_1=\frac{1}{2}k_2.\z_3J_1+\frac{1}{2}k_1.\z_3J_2\,;\,\frac{1}{2k_1.k_2}k_1.F_2.k_3J_1=\frac{1}{2}k_3.\z_2J_1-\frac{1}{2}k_1.\z_2J_3\labell{rel2}
\eeqa
 as well as  similar relations for the replacements $(1\leftrightarrow 2)$ and $(2\leftrightarrow 1,3\leftrightarrow 2)$. 

The amplitude \reef{hAAA} has been written in terms of gauge field strength so one can now write it in linear S-dual form.  
Using the doublets \reef{F2}, \reef{3} and matrix \reef{M}, one finds the following singlet:
\beqa
(\cF_1^T)_a{}^c\cM_0\cM^{23}_{bc}=e^{-\phi_0}[(*F_1)_a{}^c(*M^{23}_{bc})+F_{1a}{}^cM^{23}_{bc}]\labell{singl}
\eeqa
 Using the identity
\beqa
\eps_{a}{}^{cde}\eps_{bc}{}^{fg}=-\eta_{ab}(\eta^{df}\eta^{eg}-\eta^{dg}\eta^{ef})+\delta_a^f(\delta_b^d\eta^{eg}-\delta_b^e\eta^{dg})-\delta_a^g(\delta_b^d\eta^{ef}-\delta_b^e\eta^{df})\labell{iden3}
\eeqa
one can write the right hand side of the singlet \reef{singl} as
\beqa
(\cF_1^T)^{ac}\cM_0\cM^{23}_{bc}=e^{-\phi_0}[-\frac{1}{2}F_1^{cd}M^{23}_{cd}\eta^{a}{}_{b}+F_1^{ac} M^{23}_{bc}+(F_1)_{bc} (M^{23})^{ac}]\labell{F12}
\eeqa
There is similar singlet for the doublet $\cM^{23c}_{ab}$. Using these singlets and using the fact that graviton is traceless, \ie $\Tr(\veps.D)=2\veps_{a}{}^a$, one can write the amplitude \reef{hAAA} in exactly the  same form as \reef{ampi}.  It is amazing that the lengthy expression \reef{am4} can be written in such a simple form dictated  by the S-duality.  In the next section we will find the dilaton amplitude and study its S-duality.

\subsection{Dilaton amplitude}

The dilaton amplitude can be read from the graviton amplitude by replacing the graviton polarization with
 \beqa
 (\veps)_{\mu\nu}&=&\eta_{\mu\nu}-\ell_{\mu}(p)_{\nu}-\ell_{\nu}(p)_{\mu}
 \eeqa
 where the auxiliary field $\ell$ satisfies $\ell.p=1$ and should be canceled in the final amplitude. The graviton amplitude \reef{hAAA} satisfies the Ward identity, \ie if one replaces the graviton polarization $(\veps_1)_{\mu\nu}$ with $\z_{\mu}(p_1)_{\nu}+\z_{\nu}(p_1)_{\mu}$ where $\z_{\mu}$ is an arbitrary vector, the amplitude becomes zero. Since we have not used the traceless condition for the graviton in this amplitude,  it is obvious that the  replacement $-\ell_{\mu}(p_1)_{\nu}-\ell_{\nu}(p_1)_{\mu}$ for the graviton polarization gives zero result. Therefore, the dilaton amplitude can be read from \reef{hAAA} by replace the graviton polarization with $\eta_{\mu\nu}$ which gives the following result:
 \beqa
A_{\phi AAA}&\!\!\!\!\!\sim\!\!\!\!\!& e^{-3\phi_0/2} \Phi F_{1}^{ab}\bigg[(F_2.F_3)_{ab}+\frac{1}{2k_1.k_3} (F_{2})_{ab}k_1.F_3.k_2-\frac{1}{2k_1.k_2} (F_{3})_{ab}k_1.F_2.k_3\bigg]J_1\nonumber\\
&&\qquad\qquad+(1\leftrightarrow 2)+(2\leftrightarrow 1,3\leftrightarrow 2)\labell{pAAA}
\eeqa
where $\Phi$ is the polarization of the dilaton which is one. This amplitude should be combined with the corresponding amplitude of the RR scalar to make an S-dual multiplet. In fact the above amplitude can be extended to the following S-dual multiplet:
\beqa
\cA&\!\!\!\!\!\sim\!\!\!\!\!& e^{-\phi_0/2}   (\cF^T_{1})^{ab}\delta\cM\bigg[\cM^{23}_{ab}+\frac{1}{2k_1.k_3} \cM^{23c}_{ab}k_{1c} -\frac{1}{2k_1.k_2} \cM_{ab}^{32c}k_{1c}\bigg]J_1\nonumber\\
&&\qquad\qquad\qquad\qquad+(1\leftrightarrow 2)+(2\leftrightarrow 1,3\leftrightarrow 2)\labell{pAAA1}
\eeqa
where $\delta\cM$ is the variation of matrix \reef{M} which includes the dilaton and the RR scalar field \cite{Garousi:2011vs}. The above S-dual multiplet has two components. One component is the dilaton amplitude \reef{pAAA}, and the other one is the S-matrix element of one RR scalar and three gauge fields which is 
 \beqa
A_{CAAA}&\!\!\!\!\!\sim\!\!\!\!\!& e^{-\phi_0/2}   (*F_{1})^{ab}C\bigg[M^{23}_{ab}+\frac{1}{2k_1.k_3} M^{23c}_{ab}k_{1c} -\frac{1}{2k_1.k_2} M_{ab}^{32c}k_{1c}\bigg]J_1\nonumber\\
&&\qquad\qquad\qquad\qquad+(1\leftrightarrow 2)+(2\leftrightarrow 1,3\leftrightarrow 2)\labell{pAAA2}
\eeqa
where $C$ is the polarization of the RR scalar which is one. The above amplitude is predicted by the proposal that the nonabelian S-matrix elements should satisfy the S-dual Ward identity. In the next section we compare the RR amplitudes predicted by the S-duality with the explicit calculations
. 
 \subsection{Ramond-Ramond amplitudes}
 
We have seen that the proposal that the S-matrix elements should satisfy the S-dual Ward identity predicts new S-matrix elements. The graviton S-matrix element \reef{ampi} is predicted by S-duality and confirmed by the explicit calculation. In this section we compare the RR amplitudes that have been  predicted by S-duality, with   explicit calculation.  The scattering amplitude of one R-R $n$-form and three  gauge bosons may be given by the following correlation function:
\beqa
A&\sim&<V_{RR}^{(-1/2,-3/2)}(\veps_1^{(n)},p_1)V_{A}^{(0)}(\z_1,k_1)V_{A}^{(0)}(\z_2,k_2)V_{A}^{(0)}(\z_3,k_3)>\labell{amp23}
\eeqa
where the RR  vertex operator is \cite{Billo:1998vr,Garousi:1996ad}
\beqa
V_{RR}^{(-1/2,-3/2)}&\!\!\!\!\!=\!\!\!\!\!&(P_-H_{1(n)}M_p)^{AB}\int d^2z_1:e^{-\phi(z_1)/2}S_A(z_1)e^{ip_1\cdot X}:e^{-3\phi(\bz_1)/2}S_B(\bz_1)e^{ip_1\cdot D\cdot  X}:
\eeqa
where  the indices $A,B,\cdots$ are the Dirac spinor indices and  $P_-=\frac{1}{2}(1-\gamma_{11})$ is the chiral projection operator which makes the calculation of the gamma matrices to be with the full $32\times 32$ Dirac matrices of the ten dimensions. 
 In the R-R vertex operator, $H_{1(n)}$ and $M_p$ are
\beqa
H_{1(n)}&=&\frac{1}{n!}\veps_{1\mu_1\cdots\mu_{n}}\gamma^{\mu_1}\cdots\gamma^{\mu_{n}}\nonumber\\
M_p&=&\frac{\pm 1}{(p+1)!} \eps_{a_0 \cdots a_p} \ga^{a_0} \cdots \ga^{a_p}
\eeqa
where $\eps$ is the volume $(p+1)$-form of the $D_p$-brane and $\veps_1$ is the polarization of the R-R form.

 Using the propagators \reef{wpro}, one can easily calculate the $X$ and $\phi$ correlators in \reef{amp23}. 
To find the correlator of $\psi$, we use the following Wick-like rule for the correlation function involving an arbitrary number of $\psi$'s and two $S$'s \cite{Liu:2001qa,Garousi:2008ge,Garousi:2010bm}:
\beqa
 &&<:S_{A}(z_1):S_{B}(\bz_1):\psi^{\mu_1}
(z_2)\cdots
\psi^{\mu_n}(z_n):>=\labell{wicklike}\\
&&\frac{1}{2^{n/2}}
\frac{(z_{1\bar{1}})^{n/2-5/4}}
{\sqrt{z_{21}z_{2\bar{1}}}\cdots\sqrt{z_{n1}z_{n\bar{1}}}}\left\{(\gamma^{\mu_n\cdots\mu_1}
C^{-1})_{AB}+\cP(z_3,z_2)\eta^{\mu_2\mu_1}(\gamma^{\mu_n\cdots\mu_3}
C^{-1})_{AB}\right.\nonumber\\
&&\left.
+\cP(z_3,z_2)\cP(z_5,z_4)\eta^{\mu_2\mu_1}\eta^{\mu_4\mu_3}
(\gamma^{\mu_n\cdots\mu_5}
C^{-1})_{AB}+\cdots\pm {\rm perms}\right\}\nonumber\eeqa where  dots mean  sum  over all possible contractions. In above equation, $\gamma^{\mu_{n}...\mu_{1}}$ is the totally antisymmetric combination of the gamma matrices and  $\cP(z_i,z_j)$ is given by the Wick-like contraction
\beqa
\cP(z_i,z_j)\eta^{\mu\nu}&=&\widehat{[\psi^{\mu}(z_i),\psi^{\nu}(z_j)]}=\eta^{\mu\nu}{\frac {z_{i1}z_{j\bar{1}}+z_{j1}z_{i\bar{1}}}{z_{ij}z_{1\bar{1}}}}\labell{ppp}
\eeqa
 Combining the gamma matrices coming from the  correlation \reef{wicklike} with the gamma matrices in the R-R vertex operator, one finds  the amplitude \reef{amp23} has the the following  trace:
 \beqa
T(n,p,m)& =&(H_{1(n)}M_p)^{AB}(\gamma^{\alpha_1\cdots \alpha_m}C^{-1})_{AB}A_{[\alpha_1\cdots \alpha_m]}\labell{relation1}\\
& =&\frac{1}{n!(p+1)!}\veps_{1\nu_1\cdots \nu_{n}}\eps_{a_0\cdots a_p}A_{[\alpha_1\cdots \alpha_m]}\Tr(\gamma^{\nu_1}\cdots \gamma^{\nu_{n}}\gamma^{a_0}\cdots\gamma^{a_p}\gamma^{\alpha_1\cdots \alpha_m})\nonumber
 \eeqa
where $A_{[\alpha_1\cdots \alpha_m]}$ is an antisymmetric combination of the momenta and  the polarization of   the gauge bosons. 
The trace \reef{relation1} can be evaluated for specific values of $n$ and $p$. Since we are going to test the amplitudes that have been found in the previous sections by S-duality, we consider only the case of $p=3$. 

To verify the amplitude \reef{pAAA2}, we have to consider $n=0$. The trace \reef{relation1} then gives $m=4$. One particular term which is easy to calculate is the one which has $\z_2.\z_3$. For this term, the $X$ correlator gives $K$, \ie \reef{K}, the $\phi$ correlator gives $(z-\bz)^{-3/4}$, and $\psi$ correlator gives 
\beqa
\z_2.\z_3C\epsilon_{abcd}F_1^{ab}k_2^ck_3^d(z-\bz)^{7/4}\frac{\cP(x_2,x_3)}{|x_1-z|^2|x_2-z|^2|x_3-z|^2 }
\eeqa
They give exactly the corresponding term in \reef{pAAA2}. For all terms, one finds the following result in the Einstein frame:
\beqa
A_{CAAA}&\!\!\!\!\sim\!\!\!\!&e^{-\phi_0/2}C\bigg[\z_2\inn\z_3k_2.(*F_1).k_3+k_2.k_3\z_2.(*F_1).\z_3+k_2.\z_3 k_3 .(*F_1).\z_2-k_3.\z_2 k_2 .(*F_1).\z_3\nonumber\\
&&+k_2.\z_3 k_2 .(*F_1).\z_2-k_3.\z_2 k_3 .(*F_1).\z_3\bigg]J_{1}+[1\leftrightarrow 2]J_2+[2\leftrightarrow 1,3\leftrightarrow 2]J_3
\eeqa
Using the relations \reef{rel2}, one finds exact agreement with the amplitude \reef{pAAA2}. 

To verify the amplitude \reef{amp7}, we have to consider $n=2$.  The trace \reef{relation1} in this case  gives $m=2$ and $m=4$.  One of the open string states in this case is the transverse scalar, so it can be contracted only with the closed string momentum. The only contribution to $m=4$ is the following:
\beqa
T(2,3,4)&=&4\veps^{a_0}{}_{b}\epsilon_{a_0a_1a_2a_3}\z_2^{[b}\z_3^{a_1}k_2^{a_2}k_3^{a_3]}
\eeqa
Writing the RR polarization as $\veps_{a_0b}=-(*\veps)_{\alpha\beta}\epsilon_{a_0b}{}^{\alpha\beta}/2$ and using the identity \reef{iden3}, one finds that the above contribution is zero. The contributions from $m=2$ terms  give exactly  the amplitude \reef{amp7}.

 {\bf Acknowledgments}:  I would like to thank Mohsen Alishahiha  for useful  discussions. This work is supported by Ferdowsi University of Mashhad under grant 2/19375-1390/08/10.



\end{document}